# Three-dimensional resistivity switching between correlated electronic states in 1T-TaS$_2$.

Damjan Svetin[1,2], Igor Vaskivskyi[1], Serguei Brazovskii[3] and Dragan Mihailovic[1,2]

[1]Jozef Stefan Institute, Jamova 39, 1000 Ljubljana, Slovenia

[2]Jozef Stefan International Postgraduate School, Jamova 39, 1000 Ljubljana, Slovenia

[3]Laboratory of Theoretical Physics and Statistical Models (LPTMS)–CNRS, UMR 8626, Université Paris-Sud, F-91405 Orsay, France

## Abstract

Recent demonstrations of controlled switching between different ordered macroscopic states by impulsive electromagnetic perturbations in complex materials have opened some fundamental questions on the mechanisms responsible for such remarkable behavior. Here we experimentally address the question of whether two-dimensional (2D) Mott physics can be responsible for unusual switching between states of different electronic order in the layered dichalcogenide 1T-TaS$_2$, or it is a result of subtle inter-layer "orbitronic" re-ordering of its helical stacking structure.  We report on the switching properties both in-plane and perpendicular to the layers by current-pulse injection, the anisotropy of electronic transport in the commensurate ground state, and relaxation properties of the switched metastable state. Contrary to recent theoretical calculations, which predict a uni-directional metal perpendicular to the layers, we observe a large resistivity in this direction, with a temperature-dependent anisotropy. Remarkably, large resistance ratios are observed in the memristive switching both in-plane (IP) and out-of-plane (OP). The relaxation dynamics of the metastable state for both IP and OP electron transport are seemingly governed by the same mesoscopic quantum re-ordering process. We conclude that 1T-TaS$_2$ shows resistance switching arising from an interplay of both IP and OP correlations.

## *Introduction*

Layered transition metal chalcogenides are attracting general interest as very versatile and multifunctional quasi-two-dimensional materials displaying competing charge density wave order (CDW), orbital order, superconductivity, and in some cases magnetic order. 1T-TaS$_2$ is of particular interest, as it is thought to satisfy the conditions for an unusual low-temperature Mott insulating state in which the electronic charge density within each layer is hexagonally modulated by a three-directional IP CDW, leading to a hexagonal array of polarons in the form of the star of David (Fig. 1a,c) defined by the Ta displacements towards the central charged Ta atom(Fig. 1a, b)[1]. The resulting structure is commensurate with the underlying lattice (Fig. 1 b, c), but spin ordering of the localized electrons at the polaron center is frustrated within the hexagonal lattice[2], and no magnetic ordering has been reported so far. The material becomes superconducting under pressure[3] or upon doping[4-6]. Mechanically it behaves as a 2-dimensional Van der Waals solid, with exfoliation properties similar to graphene, which as first sight suggests the inter-layer coupling to be small. The material has has recently attracted further attention because it was shown to exhibit sub-35 fs photo-induced[7] resistance switching to a hidden (H) CDW state, with similar behavior induced by 40 ps electrical pulse-injection[8]. Recent reports of gate-tunable state switching to a supercooled state [9,10] and dynamical resistance switching[11] are also indicative of the existence of multiple competing orders at low temperature. In the H state, the relaxation properties[12] are strongly influenced by IP strain[13] and sample thickness[9] implying a strong susceptibility to external perturbations.

Currently, the mechanism for switching and the nature of the hidden state are hotly debated. This is closely linked to the controversial nature of the low-temperature commensurate ground state itself. In particular, the question is whether the physics is confined to the individual $TaS_2$ layers, or three-dimensional stacking plays an important role, whereby orbitronic ordering along the crystallographic c axis perpendicular to the layers is responsible for the switching and ordering [14,15]. Pulsing the voltage by the tip of a scanning tunneling microscope (STM) placed above the surface revealed the formation of domain walls in the a-b plane surrounding patches of a metallic phase distinct from the surrounding (insulating) C state[16,17]. In thin layers, the C transition is suppressed, and the resistivity approximately follows a straight line extrapolation of the nearly commensurate (NC) state resistivity to low temperatures, attributed to a supercooled NC phase[9]. Remarkably, recent theoretical calculations of the band structure of the C state [14,15,18-21] consistently predict insulating IP, and metallic OP electronic transport. The latter is attributed to a highly dispersing band crossing the Fermi level along the $\Gamma - A$ direction. This unusual duality in the predicted electronic transport relies upon the overlap of $z^2$ orbitals of Ta atoms on adjacent planes. These have a profound effect for transport, but not for the mechanical properties, which are a result of weak bonding by orbitals far from the Fermi level.

Until now, the theoretically predicted unusual anisotropy in transport could not be confirmed experimentally because of the lack of data of the interplane electronic resistivity measurements. In an attempt to resolve these questions, here we primarily address current-induced OP resistance switching and time-relaxation properties of the metastable hidden state. We also report on the OP resistivity and compare it with the IP measurements. The results are of fundamental importance for understanding the switching mechanism and also the c-axis correlations in the C ground state. On the practical side they open the possibility of thin film memristor devices in cross-bar geometry on polycrystalline thin films or c-axis devices useful for ultrafast low-power low-temperature memories.

*Experimental results*

**c-axis resistance in the C state.** While the IP resistivity $R_\parallel$ is easily measured by the conventional 4-probe method (see Methods), the c-axis resistance $R_\perp$ of 100 nm thick flakes cannot be measured this way. However, we can obtain $R_\perp$ by combining 2-probe and 4-probe measurements on either face of the same flake with a multiple contact device shown in Fig. 2a and b (see Methods for details). Fig. 2 c shows $R_\parallel$ and $R_\perp$ of a 75 nm thick flake of 1T-TaS$_2$. We note that the temperature-dependence of $\rho_\perp$ is qualitatively similar to $\rho_\parallel$. The anisotropy $\left(\frac{\rho_\perp}{\rho_\parallel}\right)_C$ shown in the insert to Fig. 2 c is similar at intermediate temperature to that previously reported [22], but shows a clear dip near the NC-C transition temperature T$_c$. It then recovers back in the NC state. Fig. 2 c also shows the c-axis resistance on a logarithmic plot, revealing a component that was usually hidden in the multicomponent sample-dependent resistivity curve[23,24]. Thermally activated behaviour of the form $\rho = \rho_0 e^{-\frac{E_A}{k_B T}}$ is seen between 40 and 140K, with similar OP and IP activation energies of $E_A = 112 \pm 10$ K and $91 \pm 10$ K respectively.

**Resistance switching.** This is performed with 2 contacts on either side of 90 $\mu$m thick flake, as shown in the insert to Fig. 3 a. The application of a 10 V, 50 $\mu$s pulse at 20 K causes switching to a persistent low-resistance state. For comparison, the IP resistance switching is shown in Fig. 3 b. Remarkably, the measured critical threshold current densities for switching are $\rho_\perp^T \sim 10^{-3} \frac{A}{\mu m^2}$ and $\rho_\parallel^T \sim 1.7 \times 10^{-3} \frac{A}{\mu m^2}$ respectively. Note that the value of the resistance after switching $R_H$ is typically different than the extrapolated NC-state resistance, implying that the H state is distinct from the supercooled NC state reached by gating of very thin flakes in a FET configuration [9]. Performing the switching experiment on the device shown in Fig. 2, qualitatively the same behavior is observed for each opposite pair of contacts. Remarkably, we observe that each of the contact pairs 1-8, 2-7, 3-6 and 4-5 as shown in Fig. 2 a, b, can be switched independently, implying that there is no cross-talk between adjacent cells. The final state 2-contact resistance as a function of applied pulse amplitude is shown in the insert to Fig. 3 b (Note that the contact resistance is not subtracted here). We see clear saturation behavior of $R_H$ with two plateaus above 1V, presumably corresponding to two different metastable states.

Fig. 5 a shows the effect of the applied pulse length $\tau_{pulse}$ on the switching. Remarkably we see that the switching is independent of pulse length up to 200 ms, in agreement with IP behavior[8]. With pulses longer than 1 s, the application of a pulse no longer causes any resistance change. Fig. 5 b shows the resistance as a function of time after long "erase" pulses as a function of $\tau_{pulse}$. For

The relaxation behavior of the H state resistance at different temperatures is summarized in Fig. 4. Above 35 K, the relaxation process is visible on a timescale of hours. The relaxation rate increases rapidly with increasing temperature, so that above 50 K it is too fast too measure with the present apparatus. During the course of the relaxation, the resistance shows discrete jumps, in agreement with previous IP relaxation[12]. Fitting the overall relaxation curve at different temperatures to a stretch exponential $R = R_0 \left(1 - Ce^{-\left(\frac{E_R}{k_B T}\right)^\beta}\right)$, we obtain $E_R = 650 \pm 100\ K$, with $\beta = 0.8 \sim 1.1$ as shown in Fig. 4 b. These values are very similar to those obtained for IP relaxation on the same sapphire substrates[12].

$\tau_{pulse} > 0.5$ s the H state reverts fully back to the C state, while the application of shorter pulses leads to intermediate metastable states with different resistances, depending on the length of the erase pulse, as indicated. This demonstrates manipulation of the system resistance state by adjustment of the pulse length, which is much more controlled than in the reverse "write" process shown in the insert to Fig. 3.

## Discussion

Before discussing the switching in detail, we first address the fundamental question of the inter-plane transport in relation to the band structure, which also has important implications for switching. The IP resistivity (Fig 2 c) has multiple components, which are difficult to assign to specific processes. The low-T upturn has been attributed to either Anderson localization[23] or a Mott gap splitting of the Ta $5d_{z^2}$ band at the Fermi level[25]. The presence of a clear gap in the density of states at the Fermi level by STM and angle resolved photoemission [26-28] apparently favors the latter interpretation, at least at low temperatures. Both the presence of both the upturn and the resistivity behavior at intermediate temperatures is sample dependent and thought to be associated with impurities or polaron fluctuations, but a specific origin for the either has not been conclusively ascertained. The data in Fig. 2 c show that the form of the $\rho_\perp(T)$ and $\rho_\parallel(T)$ curves is very similar, so whatever mechanism is controlling the behavior, is controlling both IP and OP transport. The fact that the Aarhenius fit is so good over the temperature interval 40-140 K suggests a new, as yet unidentified energy scale, which is unlikely to come from random defects. These would be expected to show a spread of activation energies. The observed activation energy $E_A$ is too small to be related to the Mott or CDW gap. One possibility is that it is related to the mobility edge, where the $E_A$ corresponds to the energy difference between the Fermi level at $E_F$ and the mobility edge $E_M$. Yet it is surprising that $E_A$ shows almost no anisotropy for IP and out-of plane hopping, which would be expected on the basis of the prediction that the OP transport is metallic, and IP behavior is insulating. At lower temperatures, the departure from the thermally activated behavoiur is consistent with the onset of variable range hopping below 40K, as previously reported [23].

To try and better understand the c-axis transport, let us consider the stacking structure. Full refinement of the structure from electron diffraction measurements by Ishiguro[29] show a chiral stacking of c-axis aligned double 1T-TaS$_2$ layers in the C state as illustrated in Fig. 1 d, with the double layers shifted relative to each other by $a_0 - b_0$, $a_0 + 2b_0$ or $-2a_0 - b_0$ (Fig. 1d). Here $a_0$ and $b_0$ are the lattice constants of the undistorted lattice defined in Fig. 1c. With such stacking, the central Ta atoms are aligned within each double layer, but not between double layers, thus breaking the central Ta orbital hybridization register in the *z* direction. However, Ischiguro et al [29] also point out that defects in the stacking order are quite common, in overall agreement with X-ray [30] and NMR data[31], and the coherence length along the c axis is $\xi_\perp \gtrsim 10\ c_0$.[15,29]. Phenomenological theory calculations of the stacking order by Nakanishi and Shiba[32] confirm that double layer stacking has a minimum energy, but also indicate that an alternative stacking configurations may exist nearby, in agreement with the observed stacking disorder in the structural data.

Ritschel at al[15] investigated the effect of two possible stackings **T**$_s$=2**a**$_C$+**c**$_C$ and **T**$_s$=**c**$_C$, where **a**$_C$ and **c**$_C$ are the distorted unit cell vectors in the C state, considering random bi-layer stacking along the three equivalent vectors (2**a**$_C$+**c**$_C$, 2**b**$_C$+**c**$_C$ and -(**a**$_C$+**b**$_C$)+**c**$_C$). Their band structure calculations for **c**$_C$ stacking predict an IP gap and metallic behavior along the *c* axis due to a single band crossing the Fermi level along the $\Gamma - A$ direction. For 2**a**$_C$+**c**$_C$ stacking, they predict metallic behavior in all three directions. Darancet et al[14] argue that the Coulomb interaction $U$ for IP hopping is smaller than the c-axis bandwidth $W_c$, suggesting that the Mott insulator picture of Fazekas and Tossatti[2] is valid for a single layer, but not for the bulk material. In this model, any insulating states in bulk material that do occur in the 1T-TaS$_2$ family of materials may be regarded as arising more from OP antiferromagnetic order than from an IP Mott localization phenomenon. All of these recent predictions would make the material a unidirectional metal conducting along the c axis, where the ubiquitous IP insulating behavior is attributed to Anderson localization[23] with a variable range hopping temperature dependence $\rho \simeq \rho_0 \exp\left(\frac{T_0}{T}\right)^{1/3}$ at low temperatures. The observed behavior presented in Fig. 2 is clearly not consistent with these predictions. Particularly problematic is the fact that the anisotropy is in the opposite direction (Fig. 2 c) to the predictions. Its T-dependence conveys some interesting information. Near the C-NC transition, the anisotropy strongly drops, showing a lag between OP and IP behavior upon warming, implying that OP coherence is lost first, followed by the IP coherence. The more slowly varying anisotropy $\frac{\rho_\perp}{\rho_\parallel} = 500 \sim 800$ in the intermediate temperature range between 40 and 140 K is evident in the region where a well defined thermally activated behavior is observed. The relatively large resistivity anisotropy is $\frac{\rho_\perp}{\rho_\parallel} = 1500 \sim 1800$ in the NC state implies an intrinsic anisotropy mechanism for the electron transport that is unrelated to the IP ordering.

**Resistance switching.** Considering that the largest change of lattice constant upon CDW re-ordering from the NC to the C state is in the c direction ($\frac{\Delta c}{c} \simeq 3 \sim 5\%$, compared with IP change $\frac{\Delta a}{a} \simeq 0.5 \sim 0.6\%$)[33]. It is tempting to attribute the switching of the resistance to some kind of c-axis re-stacking. Following the calculations by Ritschel et a[15], two metastable stackings of the orbitally ordered layers allow manipulation of salient features of the band structure, promoting the concept of controlling the properties of materials by using layer-stacking "orbitronics". Indeed, competing stacking configurations were already indicated by the Ginzburg-Landau theory calculations of Nakanishi and Shiba[32]. Darancet et al[14], who made spin-unrestricted band calculations attribute the sensitivity of the metal-insulator phase boundary to the nature of the inter-plane *magnetic* ordering between the localized spin ½ electrons at the center of each polaron. This implies that switching in 1T-TaS$_2$ may be arising more from OP antiferromagnetic - or even possibly ferromagnetic - reordering. So far the common feature of these theoretical studies which consider single or double layer physics is that metastabiliy in 1T-TaS$_2$ is a consequence of »orbitronic« or stacking order, while IP order is robust and unperturbed.

These notions are directly challenged by the observed anisotropy of the resistivity in Fig. 2. Moreover, recent STM studies of electrically switched 1T-TaS$_2$ [8,16,17], clearly show the appearance of IP reordering and the appearance of domain walls (DW). The new IP textured state is very similar to the one originally predicted [7] and is distinct from other states normally observed in equilibrium. Moreover, *both* the DWs, as well as the patches enclosed by the DWs are metallic, consistent with the conversion from polarons to itinerant states as proposed by Stojchevska et al. [7]. The threshold current densities are quite similar for c-axis and IP switching: $\rho_c \sim 10^{-3} \frac{A}{\mu m^2}$ and $\rho_{ab} \sim 1.7 \times 10^{-3} \frac{A}{\mu m^2}$ respectively, while the applied threshold electric fields are vastly different for IP and OP switching ($E_\parallel^T \sim 1\ V/\mu m$ vs. $E_\perp^T \sim 100\ V/\mu m$). These observations are consistent with the original idea that the switching is driven by charge injection [8].

Support for the notion that both IP DW formation *and* re-stacking are present comes from thermally activated relaxation of the H state in Fig. 4 shows nearly identical behaviour and a similar activation energy as previously reported for IP resistance[12]. This unambiguously implies that the IP and OP relaxation is governed by the same underlying processes. While the stretch exponential fit to the time-evolution of the resistivity relaxation should be considered to be an approximation for describing the Ostwald ripenening nucleated growth process [34,35], the steps in the relaxation process are ascribed to the relaxation of complete rows of polarons at a time as recently observed by STM[8].

The scenario is consistent with the topological protection mechanism which was invoked to explain the stability of the H state at low temperatures[7]. The two processes together explain the observed smooth relaxation superimposed on quasi-random steps shown in Fig. 3a.

Direct support for interlayer restacking after switching comes from STM image analysis by Ma et al [16], which shows that the switched phase is accompanied by well-defined IP phase shifts of the CDW order parameter in the topmost layer, and by a phase shift of the CDW order parameter relative to the layer underneath.

## *Conclusions*

Memristive behavior in oxides and chalcogenide glasses[36-39] is typically filamentary, associated with the manipulation of ions by the current into metastable low-resistance positions. In 1T-TaS$_2$ the electronic mechanism is fundamentally new and different, arising from CDW switching between different charge-ordered electronic states. The fact that 1T-TaS$_2$ shows switching behavior both IP and OP with not too different resistance ratios implies that some kind of c-axis restacking must occur: if the planes were completely independent of each other, there would be no change in the c-axis resistance after switching. So far phenomenological theory[32,40] describes the 3D electronic ordering and IP switching[7] with surprising success, including the prediction of the 1$^{st}$ order character of the phase transitions, something which so far eludes other more recent mesoscopic models. We conclude that the observed behaviour cannot be attributed solely to either IP ordering or OP double layer re-stacking, but indicates the existence of long-range-ordered 3D states in a well-defined energy minima. Recall that while no band structure calculations are able to reproduce the Mott insulator state, the appearance of the strongly insulating behavior both IP and OP signifies that the emergence of the Mott state might require more sophisticated treatment than the DMFT-like calculations performed so far.

The fact that adjacent cells in such a device shown in Fig. 1 can be switched independently implies that practically useful thin-film devices with unoriented films may be constructed in either cross-bar or lateral stripline geometry, opening flexible design options for new ultrafast low-temperature memristive memory devices based on CDW state switching.

*Methods*

The samples were synthesized using the vapor phase transport method[23]. The devices are shown in Figs. 1a, b, and Fig. 2a. Their thicknesses were measured by atomic force microscopy and are 75 and 90 nm respectively. Thin flakes of single crystal 1T-TaS$_2$ were placed over an array of contacts previously deposited on a sapphire substrate. To prevent IP current paths from the sides of the sample which might introduce an error in the measurement of the resistance perpendicular to the layers, an S8 polymer mask was then deposited at the edges of each contact as indicated. Finally, top gold contacts were made over the structure.

We cannot assume that the contact resistance is T-independent, or that the current paths are uniform between the contacts, so we need to subtract the relevant contact resistances at all temperatures without approximations. This can be done by placing 4 contacts on each side of the sample, and using a combination of 4-probe and 2-probe measurements as described in the following. Referring to Fig. 1, we name the contact resistances $R_1, R_2, ... R_8$, and the bare *sample* resistances between the contacts as $R^s_{2-3}, R^s_{6-7}, R^s_{2-7}, R^s_{3-6}$. The standard 4-probe measurement using contacts 1-2-3-4 and 5-6-7-8, sourcing the current between 1 and 2 or 5 and 8, and measuring the voltage between 2 and 3 or 6 and 7 respectively, gives us $R^s_{2-3}$ and $R^s_{6-7}$ respectively. The 2-probe measurements between $R_{2-3}$ and $R_{6-7}$ include the contact resistances such that $R_{2-3} = R^s_{2-3} + R_2 + R_3$, $R_{6-7} = R^s_{6-7} + R_6 + R_7$, $R_{2-7} = R^s_{2-7} + R_2 + R_7$ and $R_{3-6} = R^s_{3-6} + R_3 + R_6$. From these four equations we can calculate the sum of the c-axis resistances $R^s_{2-7} + R^s_{3-6} =$ $R_{2-7} + R_{3-6} - [(R_{6-7} - R_{6-7}^s) - (R_{2-3} - R_{2-3}^s)]$. Assuming uniform sample thickness $t$ and equal contact area $A$, the OP resistivity is then given by $\rho_\perp = (R_{2-7}^s + R_{3-6}^s)A/2t$ as plotted in Fig. 1.c.

## *Acknowledgments*

We wish to acknowledge fruitful discussions with Peter Prelovsek, Jernej Mravlje and Rok Žitko. We also thank Petra Sutar for supplying the samples. DM wishes to acknowledge funding from the ERC AdG 320602 "Trajectory", and SB acknowledges the funding from the NUST "MISiS" International Grant К3-2015-055.

*Figure captions*

**Figure 1.** The charge density wave polaronic structure and interplane stacking in the commensurate phase of 1T-TaS$_2$. a) Each layer supports its own CDW, which may be viewed as a hexagonal array of polarons. The Ta atom at the centre of each polaron is colored blue, while the displaced Ta atoms are red. The extent of each polaron is schematically indicated by the blue line. b) the detailed structure of the polaron showing atomic displacements. c) definition of the lattice and the CDW basis vectors within the commensurate in-plane structure and the lattice positions defining the stacking order according to [29] d) the C state helical layer stacking in the sequence 0-11-0-7-0-8-0... observed by Ishiguro and Sato [29].

**Figure 2.** The in-plane and out-of-plane resistivity of 1T-TaS$_2$ as a function of temperature measured on a multiple contact device. (a) A schematic diagram of the circuit, (b) a microscope image of the sample showing the contact configuration. The contact spacing is 10 $\mu m$. The masks (indicated) prevent unwanted currents through the side of the sample. (c) The in-plane resistance $\rho_\parallel$ is a 4-probe measurement, while $\rho_\perp$ is a 2-contact measurement, with the contact resistance subtracted (see Methods). The right panel shows a Arrhenius plot of $\rho_\parallel$ and $\rho_\perp$ revealing activated behavior over more than one order of magnitude of resistance between 40 K and 140K. The insert shows the anisotropy of the resistivity $\rho_\parallel/\rho_\perp$ in the C state.

**Figure 3.** Resistance switching in 1T-TaS$_2$ at 20 K. a) the temperature dependence of the c-axis resistance (including contacts) before and after the application of a 10 V 50 μs pulse. The insert shows the contact configuration and an image of the sample. The width of the contacts is 10 $\mu$m. Note the mask at the edges, designed to prevent current leakage from the sides. b) The switching of the in-plane resistance. The linearly extrapolated NC resistance value is shown for reference. The insert shows the threshold for switching along the c axis with 100 ns pulses.

**Figure 4.** Relaxation of the H state resistance at different temperatures after switching with a 50 $\mu s$ pulse (a). Note the steps superimposed on the smooth relaxation.  B) An Arrhenius plot of the activation energy $E_A$ obtained from the fit to the resistivity relaxation data.

**Figure 5.** Resistance switching with different applied pulse lengths (a). The C state is reset each time by heating the sample to 300 K in between measurements. (b) The relaxation of the resistance after application of "erase" pulses of different lengths (indicated), leading to different final states. (The contact resistance is not subtracted.)

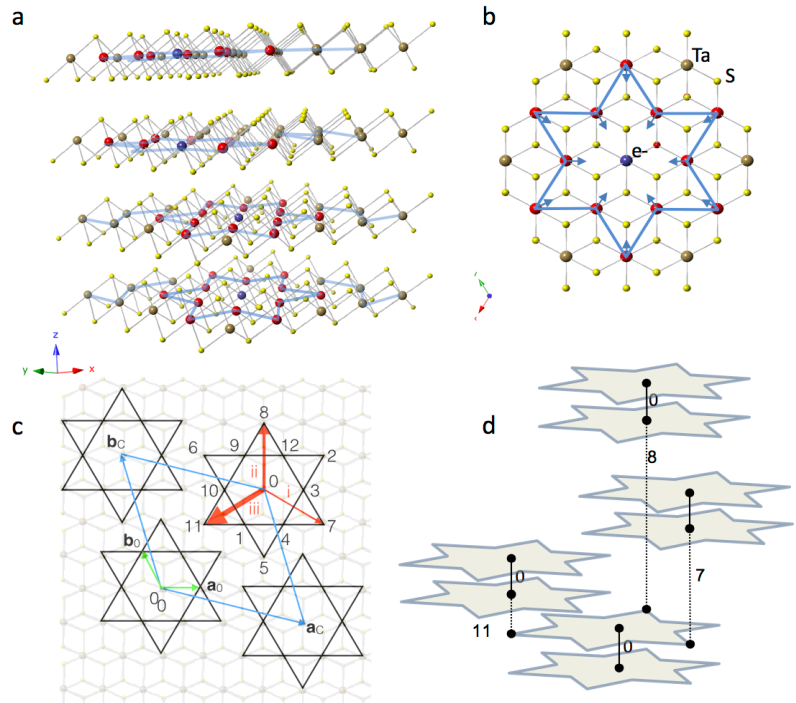

Fig. 1

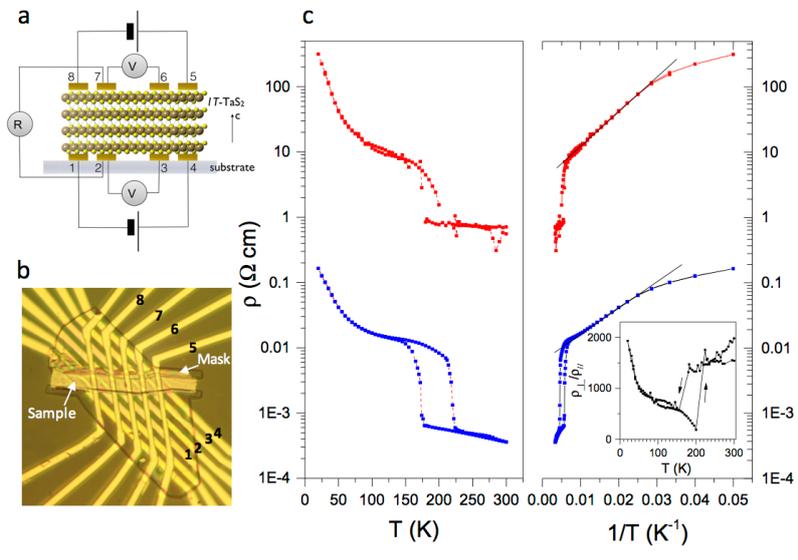

Fig. 2

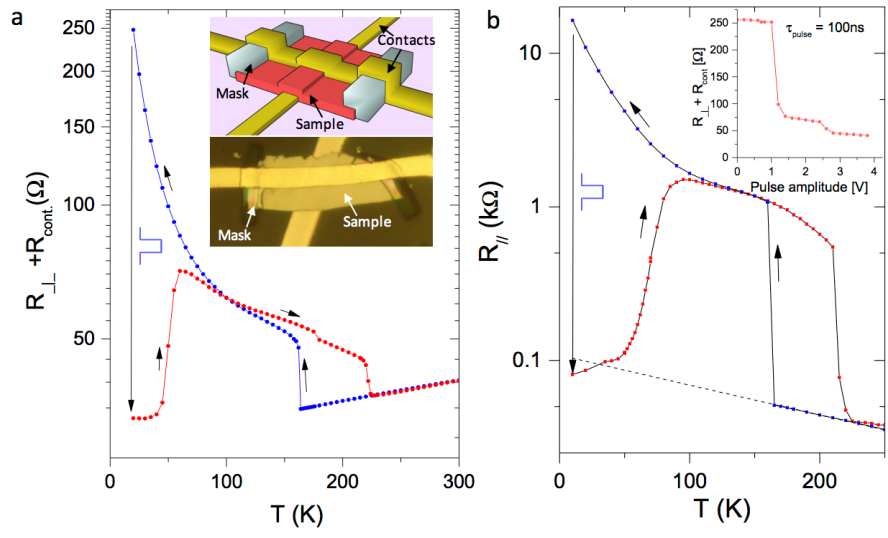

Fig. 3

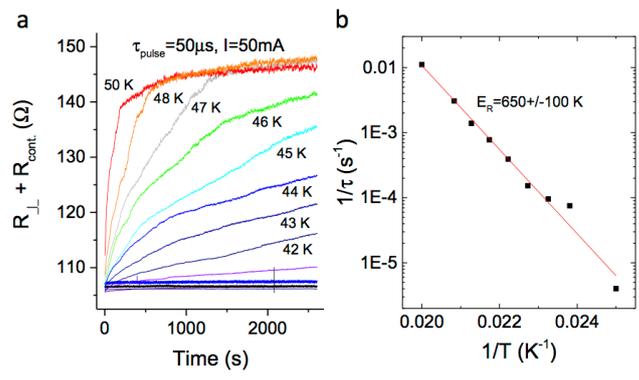

Fig. 4

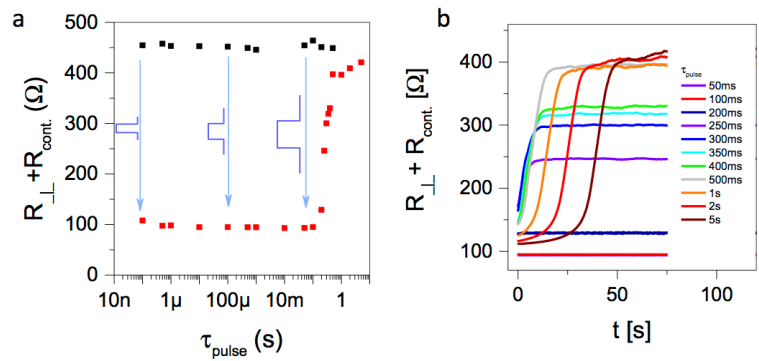

Fig. 5